\documentclass{aa}  

\usepackage{graphicx}
\usepackage{txfonts}
\usepackage{booktabs}
\usepackage{hyperref}
\hypersetup{colorlinks=true, linkcolor=blue, filecolor=magenta, urlcolor=cyan, citecolor=blue}

\newcommand{\NHI}{\ensuremath{N_\textrm{H~I}}}
\newcommand{\NNaI}{\ensuremath{N_\textrm{Na~I}}}
\newcommand{\NMgI}{\ensuremath{N_\textrm{Mg~I}}}
\newcommand{\NMgII}{\ensuremath{N_\textrm{Mg~II}}}
\newcommand{\NFeII}{\ensuremath{N_\textrm{Fe~II}}}

\begin{document}

   \title{Empirical Calibration of Na~I~D and Other Absorption Lines \\
   as Tracers of High-Redshift Neutral Outflows}

   \author{Lorenzo Moretti \inst{1}, 
          Sirio Belli \inst{1},
          Gwen C. Rudie \inst{2},
          Andrew B. Newman \inst{2},
          Minjung Park \inst{3},
          Amir H. Khoram \inst{1,4},
          Nima Chartab \inst{5},
          Darko Donevski \inst{6, 7}
          }

   \institute{Dipartimento di Fisica e Astronomia, Università di Bologna,                 Bologna, Italy
        \and
             The Observatories of the Carnegie Institution for Science, Pasadena, CA, USA
        \and
             Center for Astrophysics | Harvard \& Smithsonian, Cambridge, MA, USA
        \and
            INAF, Astrophysics and Space Science Observatory, Bologna, Italy
        \and
            Caltech/IPAC, Pasadena, CA, USA
        \and
            National Centre for Nuclear Research, Warsaw, Poland
        \and
            SISSA, Trieste, Italy
             }
             

 
  \abstract
   {
   Recent JWST observations of massive galaxies at $z>2$ have detected blueshifted absorption in Na~I~D and other resonant absorption lines, indicative of strong gas outflows in the neutral phase.  However, the measured mass outflow rates are highly uncertain because JWST observations can only probe the column density of trace elements such as sodium, while most of the gas is in the form of hydrogen. The conversion between the column density of sodium and that of hydrogen is based on observations of gas clouds within the Milky Way, and has not been directly tested for massive galaxies at high redshift.
   In order to test this conversion, we study a unique system consisting of a massive quiescent galaxy (J1439B) at $z = 2.4189$ located at a projected distance of 38 physical kpc from the bright background quasar QSO J1439+1117. The neutral outflow from the galaxy is observed as a sub-damped Lyman-$\alpha$ absorber (sub-DLA) in the spectrum of the background quasar, which enables a direct measurement of the hydrogen column density from Lyman transitions. We obtain new near-infrared spectroscopy with Magellan/FIRE and detect Na~I~D and other resonant absorption lines from Mg~II, Mg~I, and Fe~II. We are thus able to derive new, empirical calibrations between the column density of trace elements and the hydrogen column density, that can be used to estimate the mass and the rate of neutral gas outflows in other massive quiescent galaxies at high redshift. The calibration we derive for Na~I is only 30\% lower than the local relation that is typically assumed at high redshift, confirming that the neutral outflows observed with JWST at $z>2$ are able to remove a large amount of gas and are thus likely to play a key role in galaxy quenching. However, using the local calibration for Mg~II yields an order-of-magnitude discrepancy compared to the empirical calibration, possibly because of variations in the dust depletion.
   }

   \authorrunning{L. Moretti} 
   \titlerunning{Empirical Calibration of Na~D as a Tracer of High-Redshift Neutral Outflows} 
   \maketitle
%

\section{Introduction}

Gas outflows play a crucial role in galaxy formation and evolution, and may be responsible for shutting down star formation in massive systems. However, direct observational evidence for a major impact of outflows on the star formation history of massive galaxies has been  difficult to obtain, partly because the multiphase nature of galaxy outflows makes it challenging to measure the total amount of mass ejected \citep{forster-schreiber20, Veilleux_2020}. This observational challenge is one of the main obstacles in the search for a direct link between AGN-driven outflows and galaxy quenching \citep[e.g.,][]{Harrison2017, cicone18}.

At $z\sim2$, when massive galaxies start entering the quenching phase in large numbers, observational studies have been able to identify the widespread presence of ionized outflows, which can be detected via strong emission lines \citep[e.g.,][]{forster-schreiber14}; however, the corresponding mass outflow rates are low, meaning that ionized outflows are unlikely to shut off star formation and cause galaxy quenching \citep[e.g.,][]{lamperti21, concas22}. On the other hand, most of the outflow mass is expected to be in the cold phase \citep{Veilleux_2020}, which is particularly difficult to detect at high redshift. Sub-millimeter observations probe cold molecular gas, but detecting outflows in normal galaxies (as opposed to quasars) with this method is challenging \citep[e.g.,][]{herrera-camus19, spilker20, barfety25}.
Alternatively, metal absorption lines in the rest-frame optical or UV can be used to trace the neutral atomic phase. This technique has been successfully used in the local and low-redshift universe \citep[e.g.,][]{Rupke2005, Tremonti2007}, but requires a high signal-to-noise ratio on the galaxy continuum emission, which until recently was out of reach for galaxies at $z > 1$ without resorting to stacking \citep{steidel10, maltby19, taylor24} or strong gravitational lensing \citep{pettini02, jafa20, man21}.

The advent of JWST has enabled a major step forward in the field, making it possible to detect metal absorption lines due to neutral atomic gas in high-redshift galaxies. The neutral gas measurements are typically obtained from the Na~I~D doublet \citep{Davies_2024, Belli_2024, Deugenio2024, sun25}, but also Ca, Mg, and Fe absorption lines have been used  \citep{liboni25, wu25, valentino25}. 
By measuring the mass and kinematics of the dense, neutral gas phase, these studies conclude that outflows observed in massive galaxies are able to remove the majority of the gas reservoir, thus leading to galaxy quenching. However, these early results are affected by strong systematics, leading to an order-of-magnitude uncertainty on the measured mass outflow rate \citep{Belli_2024}. The major source of uncertainty comes from the fact that observations can only probe a trace element such as sodium, while most of the mass is in the form of hydrogen. A conversion of the observed column density $N_\text{X,i}$ (i.e., of element X in the i-th ionization stage) to the column density of neutral hydrogen $N_\text{H~I}$ is thus needed.
The conversion can be written as \citep[e.g.][]{Rupke2005}:
\begin{equation}
    \label{eq:one}
    N_\text{H~I} = \frac{N_\text{X,i}}{ (n_\text{X,i}/n_\text{X}) \cdot \, 10^{[\text{X/H}]} \, (n_\text{X}/n_\text{H})_\odot \, 10^{\delta_\text{X}}} \, .
\end{equation}
This calibration requires knowledge of three different physical quantities: the ionization fraction $(n_\text{X,i}/n_\text{X})$, the depletion onto dust $\delta_\mathrm{X}$, and the elemental abundance in the gas, which is written as the product of the abundance relative to solar, $10^{[\text{X/H}]}$, by the abundance in the Sun, $(n_\text{X}/n_\text{H})_\odot$.
These properties can be accurately measured in regions within the Milky Way, and the local values are adopted when studying extragalactic systems. For example, common assumptions for Na~I are $(n_\text{Na~I}/n_\text{Na}) = 0.1$ (meaning that only $10\%$ of the sodium is in the neutral phase) and $ \delta_\text{Na} = -0.95 $, together with $\log (n_\text{Na}/n_\text{H})_\odot = -5.69$ \citep{savage1996, Rupke2005}. The abundance of sodium in the outflow is rarely measured, and for massive galaxies it is often assumed that $[\text{Na/H}] = 0$, i.e. the gas has solar abundance. With these assumptions, using Eq.~\ref{eq:one} the Na-to-H conversion becomes
\begin{equation}
    \label{eq:two}
    \log N_\text{H~I} = \log N_\text{Na~I} + 7.64 \,.
\end{equation}

There are several limitations with the calibration discussed above. First, variations in dust depletion, ionization correction, and metal abundance along different lines of sight in the Milky Way can lead to differences in the calibration by up to an order of magnitude \citep{rupke02}. Second, it is expected that the physical conditions in large-scale outflows at high redshift are substantially different compared to those in the Milky Way due to, for example, the presence of shocks, photoionization by AGN, and potentially different dust composition.

In the present study we address these limitations by directly measuring the column density for both sodium and hydrogen in the large-scale outflow near J1439B, a massive quiescent galaxy at $z=2.4189$ \citep{Rudie2017}. This is possible because of the chance alignment between the galaxy outflow and a background quasar, which emits a strong, continuous spectrum extending into the rest-frame UV. 
Background quasars have been successfully used to probe outflows in star-forming galaxies at $z<1$ \citep[e.g.,][]{bouche12, Schroetter19, Martin19}; however, J1439B represents the only case where the background quasar probes the outflow for a high-redshift quiescent galaxy, and it is thus particularly suited for the study of galaxy quenching at $z\gtrsim2$.

In this work, we combine literature measurements of the H~I column density in the outflow derived from high-resolution optical spectroscopy \citep{Srianand2008, Noterdaeme2008} with new measurements of Na~I~D and other resonant absorption lines obtained from Magellan/FIRE near-infrared spectroscopy.
We thus provide an empirical, direct calibration of Na~I, Mg~I, Mg~II, and Fe~II as tracers of hydrogen in neutral outflows, which is independent of the assumptions needed by previous methods.
In Section~\ref{sec:system} we review the physical properties of the system known from previous studies, and perform a fit to the broadband photometry of the galaxy to estimate its properties. In Section~\ref{sec:spectroscopy} we introduce new near-infrared spectroscopic data and describe the equivalent width measurements. In Section~\ref{sec:results} we provide empirical calibrations for Na~I, Mg~II, Mg~I, and Fe~II. Finally, we summarize and discuss our results in Section~\ref{sec:summary}.

\section{The System}
\label{sec:system}

The unique system studied in this work consists of J1439B, a massive galaxy at $z=2.4189$, and a sub-damped Lyman-$\alpha$ absorber (sub-DLA) at roughly the same redshift, seen in the spectrum of the background quasar QSO J1439+1117. The separation between the galaxy and the QSO line of sight is $4\farcs7$, corresponding to 38 physical kpc at the galaxy redshift.

\begin{figure}[tbp]
    \centering    \includegraphics[width=0.92\columnwidth]{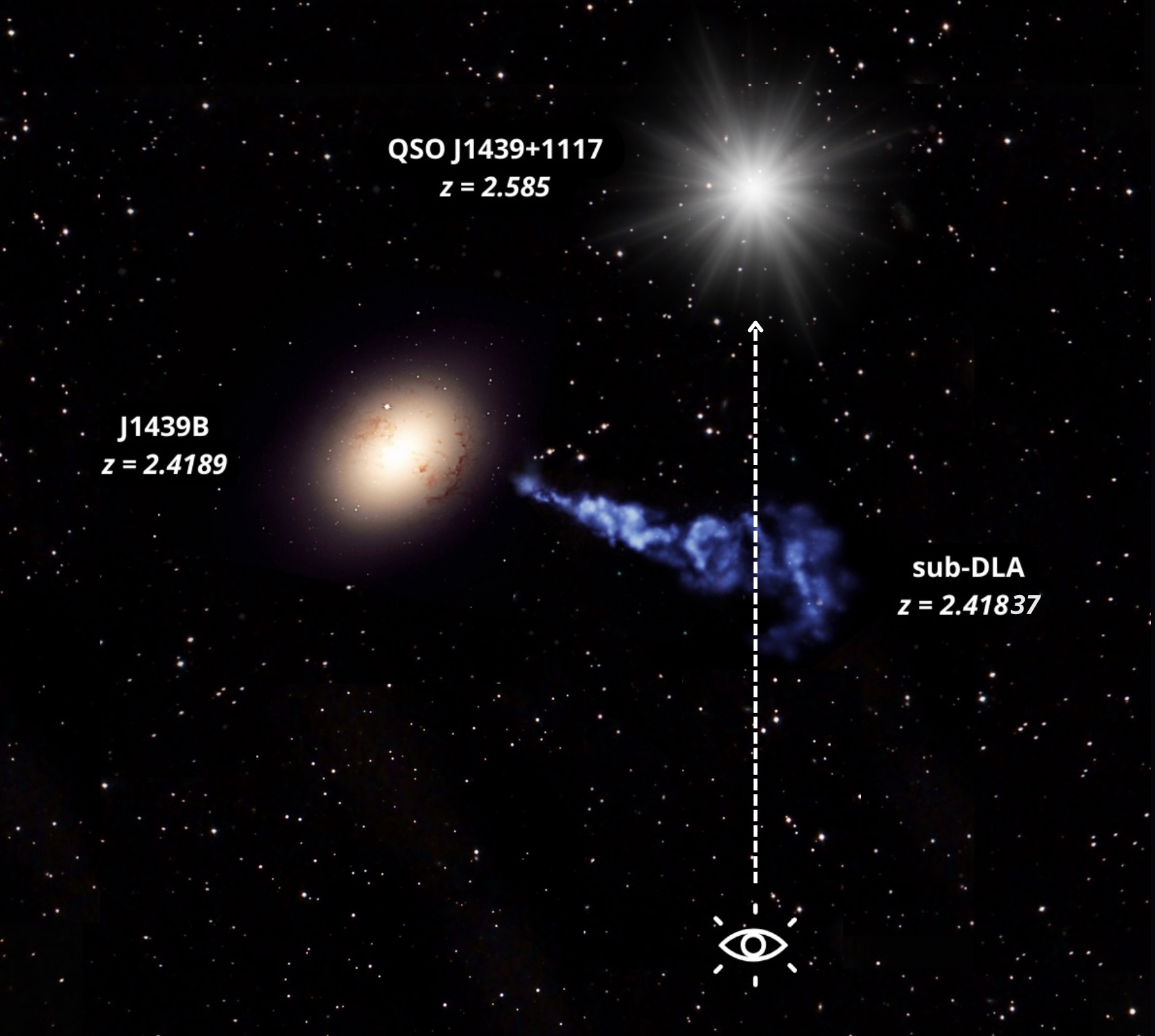}
    \caption{Artistic illustration of the system, consisting of the galaxy J1439B and its outflow, detected as a sub-DLA in the spectrum of the background quasar J1439+1117.}
    \label{fig:illustration}
\end{figure}

We interpret the sub-DLA as part of a large-scale outflow ejected by J1439B, as illustrated in Figure~\ref{fig:illustration}. This interpretation is motivated by the peculiarly high metallicity and molecular gas fraction measured in the sub-DLA, which indicate that the gas must have been ejected by a massive galaxy. Moreover, the velocities of the sub-DLA components span a range of about 1,000~km/s, pointing to an energetic process and ruling out an inflow origin for the observed gas, since inflow velocities are typically much smaller \citep[e.g.,][]{goerdt15, nelson16}. In this section we summarize the properties of both the galaxy and the outflow; see \citet{Rudie2017} for a more thorough analysis of the system.

\subsection{The Outflow}

The outflow of J1439B was first discovered as a $z=2.41837$ absorber in the spectrum of QSO J1439+1117 by \citet{Srianand2008} and \citet{Noterdaeme2008}, who characterized its physical properties using a high-resolution VLT/UVES spectrum. Analysis of the Ly-$\alpha$ damping wings and of the Ly-$\beta$ and Ly-$\gamma$ absorption lines yield a column density of neutral hydrogen of $\log N_\text{H~I} = 20.1 \pm 0.1$, which is just short of the classical DLA threshold ($\log N_\mathrm{HI} > 20.3$). Four additional components are detected, with velocities in the range 0 to $-1000$~km/s and column densities a factor of at least 5 smaller compared to the main absorber.

What makes this sub-DLA particularly interesting is its high chemical enrichment, with a roughly solar metallicity, and the detection of molecular gas through CO absorption lines, with a molecular fraction $ f = 2 N_\mathrm{H_2} / (N_\mathrm{H~I} +  2 N_\mathrm{H_2} ) = 0.27 \pm 0.10 $ \citep{Srianand2008}. These properties are highly unusual for a DLA or sub-DLA system, but are in line with expectations for gas outflowing from a massive, evolved galaxy.

\subsection{The Galaxy}
\label{sec:galaxy}

The galaxy J1439B was discovered and analyzed by \citet{Rudie2017} when looking for possible counterparts to the sub-DLA at $z=2.41837$. A near-infrared spectrum of the galaxy, obtained with Magellan/FIRE, reveals H$\alpha$, H$\beta$, [N~II] and [O~III] emission lines at a redshift $z=2.4189$, corresponding to a separation from the main sub-DLA component of 47~km/s along the line-of-sight. The line ratios are inconsistent with ionization by young stars, and lie in the AGN region of the BPT diagram \citep{baldwin81}. Moreover, the relatively large velocity dispersion in the forbidden line [O~III] suggests the presence of an ionized outflow, likely driven by AGN activity.

\citet{Rudie2017} also obtained broad-band photometry of J1439B in seven optical and near-infrared bands using the Magellan telescopes. Fitting the spectral energy distribution (SED) with FAST \citep{Kriek_2009} under the assumption of an exponentially declining star formation history yields a young, massive galaxy ($\log M_\ast/M_\odot = 10.7$) with relatively low star formation rate (SFR $= 7^{+17}_{-4} \, M_\odot \, \mathrm{yr}^{-1}$), which places J1439B about 0.7~dex below the main sequence of star formation \citep[e.g.,][]{Shivaei2015, leja22}.

\subsection{\texttt{Prospector} Fit to the Broadband Photometry}

In order to further constrain the physical properties of galaxy J1439B, we perform a new SED fit of the photometric measurements provided by \citeauthor{Rudie2017}.
We use the \texttt{Prospector} code \citep{Johnson2021}, which employs a substantially more flexible model compared to FAST, thus leading to more accurate results and realistic uncertainties.

The galaxy emission is modeled using a non-parametric star formation history with 14 time bins, adopting a continuity prior to ensure smooth transitions between adjacent time bins \citep{Leja2019}. The stellar templates are taken from the Flexible Stellar Population Synthesis \citep[FSPS,][]{conroy09, conroy10}, assuming a \citet{chabrier03} initial mass function (IMF), with the redshift fixed to the spectroscopic value.
The model also includes parameters for metallicity and dust attenuation, which is treated using the parametrization of \cite{Kriek_2013}. The attenuation $A_V$ and the ratio of diffuse-to-birth-cloud dust are left free, while the dust extinction index, which quantifies deviations from the canonical \citet{calzetti00} extinction law, was fixed to 0. This choice is supported by a recent study of massive quiescent galaxies with deep JWST spectroscopy \citep{park2024}.
Contributions from dust emission and nebular emission are not included in the model.

\begin{figure}[tbp]
    \centering
    \includegraphics[width=\columnwidth]{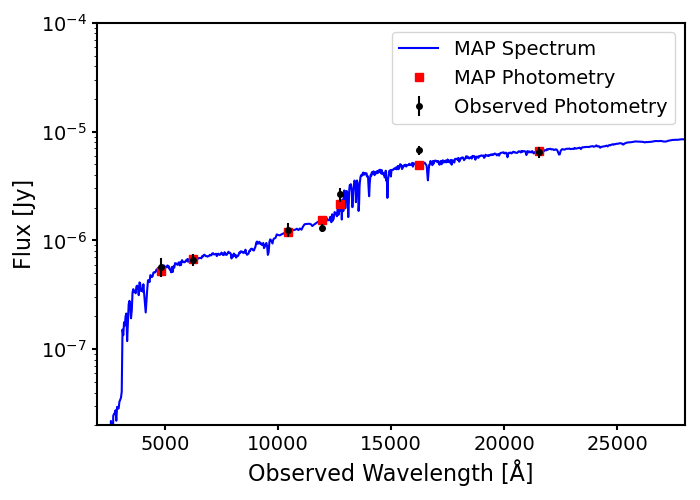}
    \caption{SED fit to the broadband photometry of galaxy J1439B. The blue line represents the model spectrum, red squares represent the best-fit model photometry, and black circles with error bars indicate the observed photometry.}
    \label{fig:SED}
\end{figure}

To account for potential systematics, we add in quadrature a $5\%$ uncertainty to the photometric error, and include the fraction of possible photometric outliers as a free parameter in \texttt{Prospector}.
The SED fitting was performed using the nested sampling method from the \texttt{dynesty} package \citep{Speagle2020}. 
The best-fit SED model, evaluated at the maximum a posteriori (MAP) point in parameter space, fits well the observed broadband photometry, as shown in Figure~\ref{fig:SED}. 

The \texttt{Prospector} fit yields $\log M_\ast/M_\odot = 10.9 \pm 0.2$, log~SFR$/(M_\odot~\mathrm{yr}^{-1}) = 1.2 \pm 0.8$, and $A_V = 0.6 \pm 0.4$. These results are in broad agreement with those obtained by \citet{Rudie2017} using FAST, but with a much larger uncertainty on the measured parameters. The best-fit star formation history presents a decline over the last Gyr, suggesting a recent quenching phase, but the posterior distribution encompasses a large range of shapes and normalizations. We conclude that the available photometry for J1439B is insufficient for a robust measurement of the star formation history. However, the quiescent nature of the galaxy is independently confirmed by the FIRE spectrum obtained by \citet{Rudie2017}: the observed H$\alpha$ luminosity, which is certainly contaminated by AGN emission, gives an upper limit of SFR$ < 13 M_\odot$/yr using the \citet{kennicutt98} calibration converted to a Chabrier IMF. The SFR estimate accounts for dust attenuation derived from the observed Balmer decrement, which is fully consistent with the best-fit value of $A_V$ obtained with \texttt{Prospector}.

As discussed in Section~\ref{sec:galaxy}, J1439B likely hosts an AGN, which may in principle affect the results of SED fitting. \texttt{Prospector} includes a model for the emission by hot dust in the AGN torus; however this is only relevant in the mid-infrared \citep{leja18}. Since we do not have observations of J1439B beyond the near-infrared, we do not include the AGN model in the \texttt{Prospector} fit.
The presence of strong AGN emission in the rest-frame UV and optical range is unlikely, given that a fit based only on stellar and dust components is able to reproduce the observed photometry reasonably well. Moreover, the stellar Balmer break is clearly detected in the observed photometry (see Figure~\ref{fig:SED}), thus confirming that the AGN continuum emission at these wavelengths is negligible compared to the stellar component. 

\begin{figure*}[t]
    \centering
    \includegraphics[width=\textwidth]{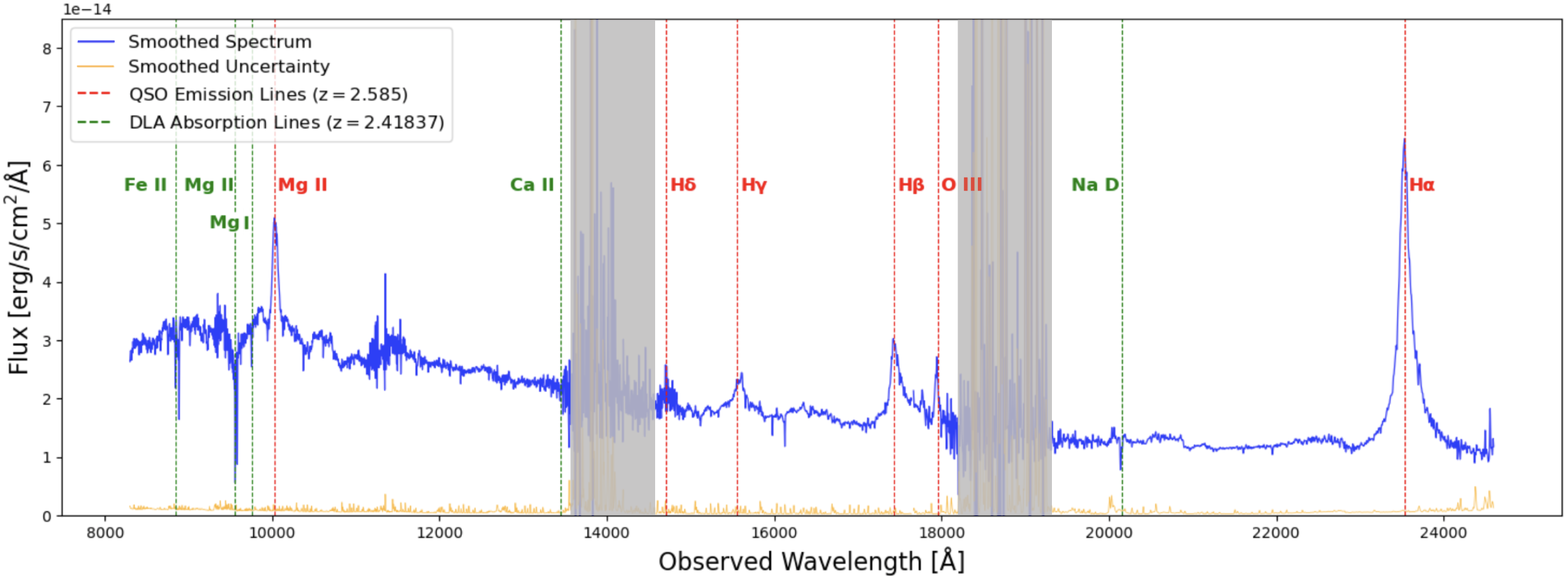}
    \caption{\textit{Magellan}/FIRE spectrum of QSO J1439+1117 (in blue; smoothed with a 4-pixel Gaussian kernel) and spectral uncertainty (in orange). Red dashed lines mark the main emission lines from the QSO at z = 2.585, while green dashed lines mark absorption lines from the sub-DLA at z = 2.41837. Spectral windows with poor atmospheric transmission are marked in gray.}
    \label{fig:spectrum}
\end{figure*}

\section{Absorption Line Measurements}
\label{sec:spectroscopy}

\subsection{Spectroscopic Data}
We observed QSO J1439+1117 with the Folded-port InfraRed Echellette (FIRE) spectrograph on the Baade \textit{Magellan} Telescope in Chile. FIRE is a near-infrared, cross-dispersed echelle spectrograph providing medium-resolution spectroscopy over the full near-infrared range, 0.8–2.5 µm. Observations were conducted on March 9–10, 2024, using a 6'' × 0.6'' slit, yielding a spectral resolution $R \sim 6{,}000$.
We adopted an AB dithering pattern with 20-minute exposures offset by 3''. The total exposure time was 160 minutes under good weather conditions, with 0.6'' seeing in the near-infrared.
The observations were reduced with the FIREHOSE pipeline \citep{firehose}, which includes flat-fielding, wavelength calibration, illumination correction, and slit tilt correction. We also observed an A0V star close to the target, and used its spectrum to perform the telluric correction using the \textit{xtellcorr} package \citep{vacca03}.
After extracting the 1D spectrum for each of the A and B nod positions, we detected a spurious absorption feature in the source spectrum at 20148.7~$\AA$, close to the Na~D doublet. This is due to a detector defect that lies near the source spectrum only in the A position. When combining the A-B and B-A spectra, we thus leave out the A-B spectrum in the region around Na D, obtaining a combined spectrum that has half the exposure time but is free from artifacts.
The final 1D spectrum is shown in Figure~\ref{fig:spectrum} and is made publicly available\footnote{\url{https://doi.org/10.5281/zenodo.15237852}}. We note the presence of both emission lines from the QSO at $z=2.585$ and absorption lines from the sub-DLA at $z=2.41837$.

This system has also been previously observed in the optical by \cite{Srianand2008} and \cite{Noterdaeme2008} using the Ultraviolet and Visual Echelle Spectrograph (UVES) on the \textit{Very Large Telescope} (VLT), under ESO program 278.A-5062(A). The UVES spectrum covers the 3300–7100~$\AA$ range with a resolving power of $R = 50{,}000$.
We use this high-resolution spectrum only for visualization purposes. We adopt the hydrogen column densities measured from the UVES data for each velocity component by \citet{Srianand2008}.

\subsection{Equivalent Width Measurements}

\begin{figure}[t]
    \centering
    \includegraphics[width=\columnwidth]{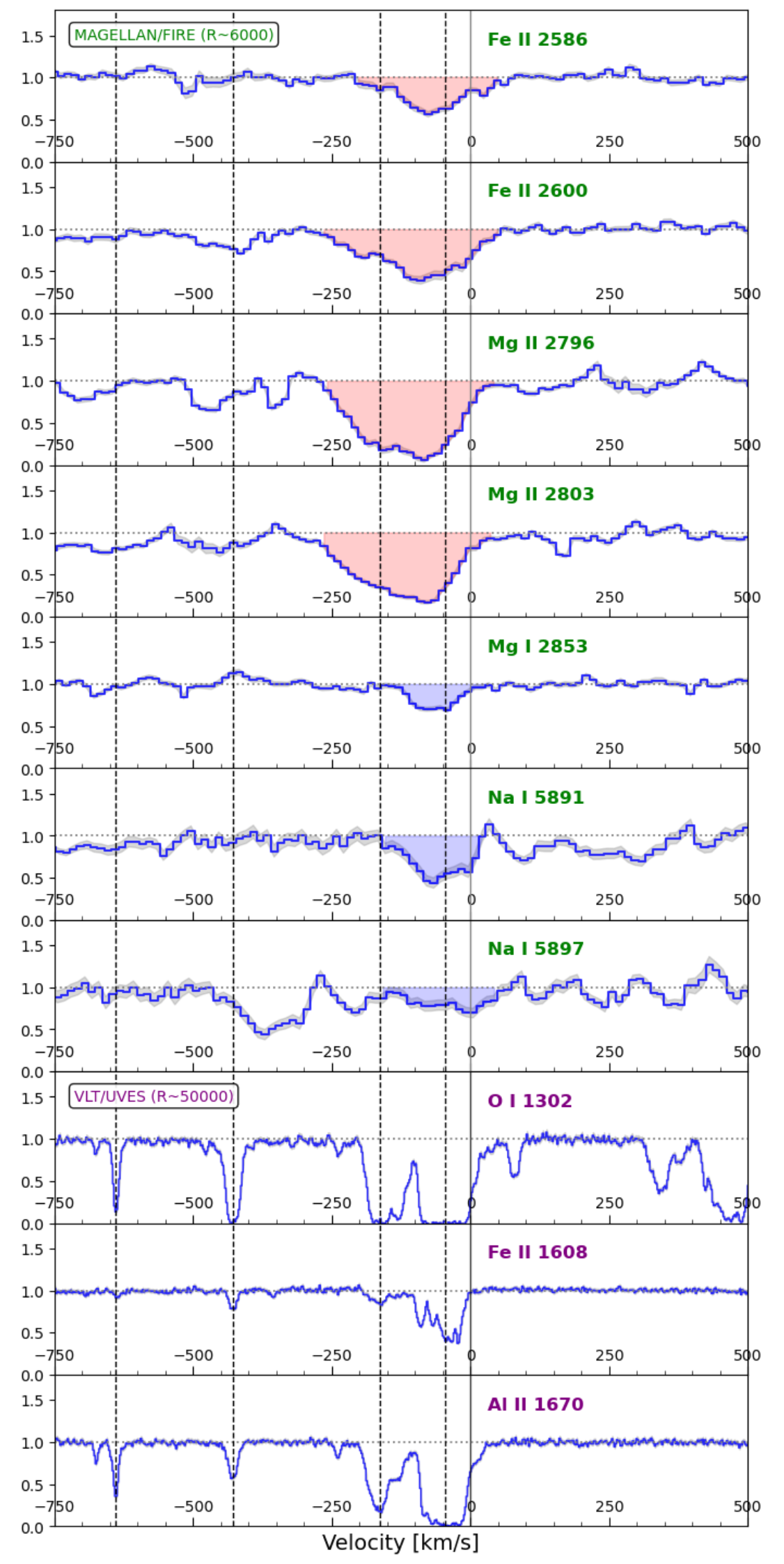}
    \caption{Absorption features in the QSO J1439+1117 spectrum in velocity space, with zero corresponding to the systemic velocity of galaxy J1439B. Normalized flux is shown on the $y$-axis of each panel. The top panels show Fe~II, Mg~II, Mg~I and Na~I transitions in the FIRE spectrum. The shaded areas represent the spectral regions used in the calculation of the EW for the component at -47~km/s (blue shaded areas), or for the sum of the components at -47 and -164~km/s when they are blended (red shaded areas).
    The bottom panels show O~I, Fe~II, and Al~II in the higher-resolution UVES data. Vertical dashed lines mark the absorbers used to fit the H~I distribution in \citeauthor{Srianand2008} and \citeauthor{Noterdaeme2008}.}
    \label{fig:stack}
\end{figure}

In the FIRE spectrum  we identify several resonant absorption lines due to cold gas: Fe II~2586, 2600, Mg~II 2796, 2803, Mg~I 2853, and Na~I 5891, 5897 (the Na~I~D doublet). These are shown in Figure~\ref{fig:stack}, plotted with respect to the systemic redshift of the J1439B galaxy, and with the spectral continuum normalized to a value of 1 in each region. We also look for the Ca~II 3934, 3969 doublet (Ca~II H and K lines), but at the sub-DLA redshift these wavelengths fall into a region with very low atmospheric transmission, making it impossible to detect any absorption. In the bottom of Figure~\ref{fig:stack} we also show a select sample of transitions in the rest-frame UV from the UVES spectrum: O~I 1302, Fe~II 1608 and Al~II 1670. Vertical dashed lines identify four velocity components for which \citeauthor{Srianand2008} and \citeauthor{Noterdaeme2008} measure the H~I column density from Voigt fits to the Lyman absorption lines. These components are found at $-47$, $-164$, $-428$, and $-640$~km/s compared to the galaxy systemic redshift, and their column densities are $\log~\NHI=$~20.1, 19.4, 19.2, and 19.2 respectively. A fifth component at $-938$~km/s is much weaker ($\log~\NHI=17.8$) and falls outside of the velocity range shown in the figure. 

The kinematic structure of the sub-DLA in the high-resolution UVES spectrum appears complex, with the outflow consisting of multiple cloudlets, each traveling at a different velocity, as indicated by the presence of distinct absorption profiles of the same element at varying velocities. All absorption lines are blueshifted by several tens to hundreds of kilometers per second — clear evidence of a neutral outflow moving toward the observer. The absorption lines in the FIRE spectrum are aligned with the strongest velocity components, but appear broader due to the lower spectral resolution. 
Moreover, for the Mg~II and Fe~II lines, saturated absorption causes blending between the main component of the sub-DLA at $-47$~km/s and the adjacent component at $-164$~km/s, similarly to what is observed for O~I in the UVES spectrum.

We measure the equivalent width EW of each transition in the FIRE spectrum in order to quantify the strength of the absorption. 
For Mg~I and Na~I we integrate the spectrum only in the wavelength region corresponding to the strongest velocity component at -47 km/s (blue shaded areas in Figure~\ref{fig:stack}); while for Mg~II and Fe~II we measure the combined EW of the -47 km/s and -164 km/s components, which are blended (red shaded areas in the figure).
All EW measurements are reported in Appendix~\ref{sec:appendix}.

\section{Empirical Calibrations of the Column Densities} 
\label{sec:results}

In this section we compare the column density of hydrogen to the column density of sodium, magnesium, and iron measured in the sub-DLA. The goal is to obtain an empirical calibration that can be used in future studies to estimate the hydrogen column density of a system given the measured column density for one of the metals.

The hydrogen column density of the sub-DLA is already available from rest-UV spectroscopy \citep{Srianand2008}. For the metal lines, we derive the column density from the EW measured in Section~\ref{sec:spectroscopy}, making use of the relation valid in the optically thin regime \citep[e.g.,][]{draine11}:  
\begin{equation}
\label{eq:ew}
    \mathrm{EW} = \frac{\pi e^2}{m_e c^2}\, \lambda_0^2 \, N f \simeq \left( 8.85\cdot10^{-13}~\textrm{cm} \right)~ \lambda_0^2 \, N f \,,
\end{equation}
where $\lambda_0$ is the restframe wavelength of the transition and $f$ is its oscillator strength. When the regime becomes optically thick, then the same observed EW corresponds to a higher number of atoms, and this relation can only provide a lower limit on the column density. The doublet ratio can be used to discriminate between the two regimes: in the optically thin case the doublet ratio is roughly equal to the ratio of the oscillator strengths, while in the optically thick case the absorption lines become saturated, and the doublet ratio becomes closer to 1. Measurements of doublet ratios and column densities are reported in Appendix~\ref{sec:appendix}.

\subsection{Sodium}

The doublet ratio for Na~I, measured for the -47 km/s velocity component, is consistent with the theoretical value of $\sim2$ and indicates that the absorption is optically thin. We can thus derive a robust measurement of the Na~I column density (see Table~\ref{tab:nai}). Comparing this to the H~I column density for the same velocity component, we obtain a column density ratio $\log (\NHI/\NNaI) = 7.5\pm0.1$. We now assume that the relation between \NHI~and \NNaI~is linear, analogously to what is done in Equation~\ref{eq:one} for the local calibration. With this assumption, we derive an empirical calibration that can be used to infer the hydrogen mass from the observed Na~I absorption in other outflows:
\begin{equation}
\label{eq:calib_na}
    \log \NHI = \log \NNaI + 7.5 \,.
\end{equation}
This calibration does not depend on the assumed values for dust depletion, ionization correction, and abundance, because it is entirely based on the observed number of Na~I and H~I atoms in the sub-DLA.
Our empirical calibration yields a hydrogen column density that is 0.14 dex, or about 30\%, smaller than the local calibration by \citet{Rupke2005} given in Eq.~\ref{eq:two}.
The discrepancy is not particularly large given that the local calibration depends on depletion, ionization correction, and abundance assumptions which have high levels of uncertainty.

\subsection{Magnesium}

The Mg~II doublet ratio, measured for the sum of the velocity components at -47 km/s and -164 km/s, is consistent with unity, indicating strong line saturation since the optically thin value is $\sim2$. We thus use the observed EW to derive a lower limit for the Mg~II column density, which we compare to the H~I column density for the two velocity components (see Table~\ref{tab:mgii}). This results in an upper limit on the Mg~II-to-H~I calibration.
To further constrain the calibration we also consider the shallower — and thus unsaturated — Mg~II absorption features at $v=-428$~km/s and $v=-640$~km/s. In these cases the absorption lines are undetected, and thus yield an upper limit on the column density and a lower limit on the calibration. We discard the $v=-428$~km/s component because the spectrum is affected by strong residuals of unknown origin. For the component at $v=-640$~km/s we measure an upper limit on the EW in the following way. We assume that the Mg~II absorption is spectrally unresolved in the FIRE data, which is confirmed by looking at the line widths of UV transitions in the UVES spectra (Figure~\ref{fig:stack}). Given the FIRE spectral resolution $R\sim6000$, we calculate a line width of 85~km/s, corresponding to 2$\sigma$ on either side. We then measure the largest fluctuation of the normalized spectrum within the expected line width, and multiply this number by the expected line width (in angstrom) to obtain the upper limit on the EW. This is a conservative measurement that accounts for systematic uncertainties and imperfect normalization of the spectrum, and if some of the observed features are due to genuine Mg~II absorption, the true EW would always be lower than our estimated upper limit. We then convert this EW limit to an upper limit on the Mg~II column density, which we compare to the H~I column density of the velocity component at -640~km/s (see Table~\ref{tab:mgii}).

Analysis of different velocity components yielded an upper and a lower limit for the Mg~II-to-H~I column density ratio. Each velocity component corresponds to a different gas clump, and we assume that these clumps have similar physical conditions, because their column densities differ by an order of magnitude at most ($19.2 < \log \NHI < 20.1$). We thus combine the upper and lower limits together, obtaining a relatively small range of values: $6.2 < \log(\NHI/\NMgII) < 6.4$. We can thus obtain an approximate empirical calibration:
\begin{equation}
\label{eq:calib_mg}
    \log \NHI = \log \NMgII + 6.3 \;.
\end{equation}
This empirical calibration yields an H~I column density that is about an order of magnitude larger compared to the local calibration (Eq.~\ref{eq:one}) with the standard assumptions, i.e. that most of the Mg atoms are singly ionized, the Mg abundance equals the solar value $\log (n_\text{Mg}/n_\text{H})_\odot = -4.4$, and the dust depletion is $\delta_\mathrm{Mg}=-0.8$ \citep{savage1996, jenkins09, wu25}. Some of the standard local assumptions must therefore be substantially incorrect.

We can directly test the degree of ionization of Mg atoms in the absorbing gas using the Mg~I transition detected in the FIRE spectrum. We derive a column density for the $v=-47$~km/s component of $\log \NMgI=12.3$ (see Table~\ref{tab:mgi}), which is at least 30 times smaller than the column density for Mg~II, since $\log \NMgII > 13.8$. A small caveat is that the Mg~II column density is measured for the blend of the $v=-47$~km/s and $v=-164$~km/s components; however nearly 90\% of neutral hydrogen is found in the $v=-47$~km/s component \citep{Srianand2008}, so that we can safely compare the Mg~I and Mg~II column densities. We conclude that most of the Mg atoms are ionized, consistent with local measurements, and that the ionization correction is not the root of the discrepancy between the high-redshift empirical calibration and the local calibration.

Finally, we also provide the Mg~I calibration derived empirically from our column density measurements:
\begin{equation}
\label{eq:calib_mgi}
    \log \NHI = \log \NMgI + 7.8 \;.
\end{equation}

\subsection{Iron}

The Fe~II 2586, 2600 doublet ratio is $2.0\pm0.1$, which is substantially lower than the optically thin value of $\sim4$ and indicates saturation. We can thus derive a lower limit on the column density of \NFeII, corresponding to an upper limit on the \NHI/\NFeII\ calibration (see table~\ref{tab:feii}).
Additionally, the UVES spectrum contains an optically thin Fe~II transition (Fe~II 1608, shown in Figure~\ref{fig:stack}), which \citet{Noterdaeme2008} use to derive a column density for the main velocity component, $\log \NFeII = 14.28 \pm 0.05$. We use this value to infer a calibration for Fe~II:
\begin{equation}
\label{eq:calib_fe}
    \log \NHI = \log \NFeII + 5.8 \;.
\end{equation}
This calibration yields a neutral hydrogen column density that is 0.4 dex lower compared to the local calibration with the default assumptions for Fe~II, i.e., all atoms are singly ionized, $\log (n_\text{Fe}/n_\text{H})_\odot = -4.5$, and $\delta_\mathrm{Fe} = -1.7$ \citep{savage1996, jenkins09, wu25}.

\begin{table*}[tb]
\centering
\renewcommand{\arraystretch}{1.3} 
\setlength{\tabcolsep}{10pt} 
\begin{tabular}{c|c|ccccc}
 & \multicolumn{1}{|p{3.5cm}|}{\centering \textbf{High-z empirical calibration}} & \multicolumn{1}{p{2.5cm}}{\centering \textbf{Local calibration}} & \multicolumn{1}{p{1.5cm}}{\centering Ionization correction} & \multicolumn{1}{p{1.5cm}}{\centering Relative abundance} & \multicolumn{1}{p{2.0cm}}{\centering Solar abundance} & \multicolumn{1}{p{2.0cm}}{\centering Dust depletion} \\
 & $\log \NHI - \log  N_\mathrm{X,i}$ & $\log \NHI - \log  N_\mathrm{X,i}$ & $n_\mathrm{X, i} / n_\mathrm{X}$ & [X/H] & $\log (n_\text{X}/n_\text{H})_\odot$ & $\delta_\mathrm{X}$ \\
\midrule
Na~I & $ 7.5 \pm 0.1 $ & $7.6$ & 0.1 & 0 & $-5.69$ & $-0.95$  \\
Mg~I & $ 7.8 \pm 0.1 $ & $(5.7, 6.7)$ & 0.1 & 0 & $-4.38$ & $(-1.27, -0.27)$ \\
Mg~II & $ 6.3 \pm 0.2 $ & $(4.7, 5.7)$ & 1 & 0 & $-4.38$ & $(-1.27, -0.27)$  \\
Fe~II & $ 5.8 \pm 0.1 $ & $(5.4, 6.7)$ & 1 & 0 & $-4.46$ & $(-2.24, -0.95)$ \\
\midrule
\end{tabular}
\vspace{2mm}
\caption{\footnotesize Comparison of our high-redshift empirical calibration to the local calibration, based on Eq.~\ref{eq:one}. The last four columns list the assumptions used to derive the local calibration. Solar abundance and dust depletion values are taken from \citet{savage1996} for Na and from \citet{jenkins09} for Mg and Fe. The ionization correction for Na~I is from \citet{Rupke2005}, while for Mg~I it is estimated from the results of \citet{murray07}.}
\label{tab:calibrations}
\end{table*}

\section{Summary and discussion}
\label{sec:summary}

We have carried out a detailed analysis of a unique system consisting of J1439B, a massive quiescent galaxy at $z=2.4189$, and a sub-DLA at $z=2.41837$ ejected by the galaxy, located $\sim38$ physical kpc away in projection. The coincidental alignment of this outflow region with a bright background quasar has allowed for a detailed study of the relative abundance of metals and hydrogen in the neutral phase of the outflowing gas.

Our main result was the derivation of empirical relations, given in Eq.~\ref{eq:calib_na}, \ref{eq:calib_mg}, \ref{eq:calib_mgi} and \ref{eq:calib_fe}, between the metal column densities derived from resonant Na~I, Mg~II, Mg~I and Fe~II absorption lines and the hydrogen column density directly measured from Lyman absorption.
Our work thus provides an empirical basis for characterizing galactic outflows in the early Universe, a key step toward quantifying the role of feedback in shaping galaxy evolution. 

It is instructive to compare our high-redshift empirical calibrations to the ones obtained by plugging local measurements into Eq.~\ref{eq:one}. We show this comparison in Figure~\ref{fig:calibration} and in Table~\ref{tab:calibrations}, where we list the commonly adopted values of ionization correction, elemental abundance, and dust depletion. 
The calibrations based on local measurements yield a hydrogen column density that is 0.14 dex and 0.4 dex higher than our empirical calibrations for Na~I and Fe~II, respectively; and 1 dex lower for Mg~II.
This means that the outflow ejected by J1439B has substantially fewer magnesium atoms than one would expect based on local observations.
Barring the existence of observational errors that we did not account for, the discrepancy suggests that high-redshift outflows are different from local clumps of neutral gas in one or more properties:

\begin{figure}[t]
    \centering
    \includegraphics[width=0.5\textwidth]{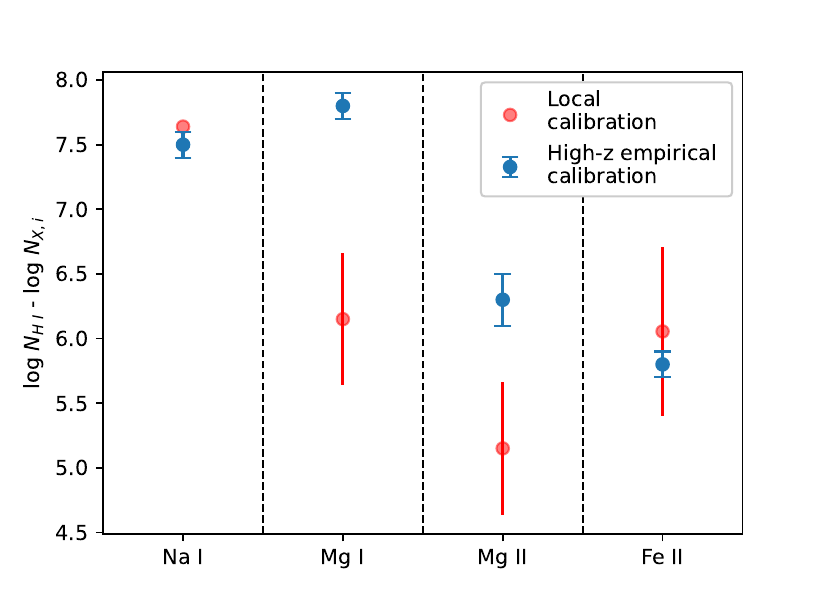}
    \caption{Column density calibration for different species. The local calibration based on Milky Way gas clouds is shown in red, with the error bar showing the range due to variations in dust depletion. Our empirical calibration derived from J1439B is shown in blue. The two calibrations yield consistent results for Na and Fe, but are inconsistent for Mg, suggesting stronger dust depletion of Mg in high-redshift outflows compared to the Milky Way.}
    \label{fig:calibration}
\end{figure}

\begin{itemize}
    \item \textbf{Ionization correction:} The ionization correction is expected to depend on the physical properties of the gas, and is particularly uncertain for Na~I, which traces a small fraction of Na atoms since they are mostly ionized \citep{murray07, baron2020}. However, for Mg we directly confirmed the theoretical expectation that Mg~II is much more abundant than Mg~I. We can thus rule out a strong impact of the ionization correction on the calibration for Mg, and also for Fe which has a similar ionization potential. 

    \item \textbf{Elemental abundances:} The gas is often assumed to have solar abundances, but this is not necessarily true since the interstellar medium of high-redshift galaxies is expected to have different chemical abundances compared to the solar neighborhood. Stellar abundance measurements for $z\sim2$ massive quiescent galaxies find a deficiency in iron,  $-0.6 < $ [Fe/H] $ < 0$, and an enhancement in [Mg/Fe], corresponding to roughly solar values of [Mg/H] \citep{beverage24, beverage25, jafa25}. In one notable case, magnesium is strongly enhanced, [Mg/H]~$\sim0.8$ \citep{jafa20}. If the gas ejected by J1439B were to be enriched in Mg, Equation~\ref{eq:one} would predict a hydrogen-to-magnesium column density ratio that would be lower than $\log \NHI/\NMgII \sim 5.3$, thus increasing the discrepancy with our empirically determined value $\log \NHI/\NMgII \sim 6.3$. It is also possible that the abundance of the gas ejected in the outflow is not necessarily representative of that in the galaxy.

    \item \textbf{Dust depletion:} Dust depletion is perhaps the most uncertain ingredient in the local calibration. In Milky Way measurements, the dust depletion of Na shows a large scatter of 0.5~dex \citep{Wakker2000}; while for Mg and Fe the scatter is smaller, but the dust depletion value depends strongly on the gas density, varying by more than 1~dex between different lines of sight \citep{jenkins09, decia16}. This also means that our assumption of a linear relation between the hydrogen and the metal column densities may not hold, if the dust depletion systematically varies with gas density and, therefore, column density. While the Fe depletion in our system seems consistent with the typical values found in the Milky way \citep[as already noted by][]{Srianand2008}, the low observed column density of Mg could be explained by a stronger depletion of Mg onto dust grains. This implies dust-to-metal ratios exceeding those observed in the Milky Way or Large Magellanic Cloud \citep{romanduval21} and in local galaxies \citep{devis19}. 
\end{itemize}

We conclude that dust depletion is the most plausible reason behind the discrepancy between the local calibration and our high-z empirical calibration. In particular, the fact that the two calibrations agree for Na and Fe but disagree for Mg is consistent with a strong Mg dust depletion due to a high dust-to-metal ratio.
Interestingly, recent theoretical work supports this scenario: using the SIMBA cosmological simulation with a realistic dust model, \citet{lorenzon25} find that about 20\% of $z\sim1-2$ quiescent galaxies experiencing strong AGN-driven feedback exhibit elevated dust-to-metal ratios compared to local analogues. Such systems may act as reservoirs of large grains that are expelled to CGM distances, where their large size ensures prolonged survival \citep{hirashita24, richie24}. Therefore, strong Mg depletion may naturally result from accretion onto preexisting grains in dense and relatively cool gas. The complexity of the physical processes at play suggests that the conditions of high-redshift outflows differ from those in the local universe.

The most important caveat in our work is that the empirical calibrations have been derived for a single line of sight in only one system. By considering the discrepancy with the local calibrations, and the range of possible values for dust depletion and the other relevant physical properties, we can conclude that our Na~I and Fe~II calibrations provide at least the correct order of magnitude for the hydrogen column density.
Clearly, it is crucial to extend this analysis to a wider sample of outflows in high-redshift massive galaxies, so that the robustness and precision of the calibrations can be assessed.

Despite the large systematic uncertainties still present in the calibrations, our measurements confirm that previous estimates of neutral gas based on resonant metal lines are likely to be correct within an order of magnitude. This is particularly important for the interpretation of recent JWST observations of massive quiescent galaxies at high redshift \citep{Belli_2024, Deugenio2024, wu25}. Our new calibrations confirm that the mass outflow rate in the neutral phase is substantially larger than that of the ionized phase, and it is sufficiently high to deplete the cold gas reservoir on rapid timescales and explain the observed quenching of star formation.

\begin{acknowledgements}
SB and AHK are supported by the ERC grant 101076080 ``RED CARDINAL''.
DD acknowledges support from the NCN through the SONATA grant UMO2020/39/D/ST9/00720 and support from the Polish National Agency for Academic Exchange (Bekker grant BPN /BEK/2024/1/00029/DEC/1).
This article includes data gathered with the \emph{Magellan} Telescopes located at Las Campanas Observatory, Chile, as well as observations collected at the European Organisation for Astronomical Research in the Southern Hemisphere under ESO program 278.A-5062(A) and obtained from the ESO Science Archive Facility.
\end{acknowledgements}

\bibliographystyle{aa}
\bibliography{references}

\begin{appendix}

\section{Measurements}
\label{sec:appendix}

Tables~\ref{tab:nai}, \ref{tab:mgii}, \ref{tab:mgi}, and \ref{tab:feii} report the EW measurements and corresponding column densities for transitions observed in the FIRE spectrum.

\begin{table}[tb]
\centering
\renewcommand{\arraystretch}{1.3} 
\setlength{\tabcolsep}{10pt} 
\begin{tabular}{c|c}
\multicolumn{2}{c}{\textbf{\large Na~I Equivalent Width and Column Density}} \\
\midrule
\midrule
\textbf{Component} & \textbf{-47 km/s} \\
\midrule
EW(Na~I 5891) & $(0.85 \pm 0.07)$ Å \\
EW(Na~I 5897) & $(0.39 \pm 0.06)$ Å \\
EW ratio & $2.2 \pm 0.4$ \\
\midrule
log(\NNaI~/~cm$^{-2}$) & $12.6 \pm 0.1$\\
log(\NHI~/~cm$^{-2}$)$^\dagger$ & $20.1 \pm 0.1$\\
log(\NHI~/~\NNaI) & $7.5 \pm 0.1$ \\
\midrule
\end{tabular}
\vspace{2mm}
\caption{\footnotesize Equivalent width and column density measurements of Na~I for the component at $v=-47$ km/s. The Na~I column density is the average of the values obtained using each of the two transitions.\\
$^\dagger$ from \citet{Srianand2008}.}
\label{tab:nai}
\end{table}

\begin{table}[tb]
\centering
\renewcommand{\arraystretch}{1.3} 
\setlength{\tabcolsep}{10pt} 
\begin{tabular}{c|c|c}
\multicolumn{3}{c}{\textbf{\large Mg~II Equivalent Widths and Column Densities}} \\
\midrule
\midrule
\textbf{Component} & \textbf{-47 km/s} \& \textbf{-164 km/s} & \textbf{-640 km/s} \\
\midrule
EW(Mg~II 2796) & $(1.77 \pm 0.02)$ Å & $< 0.54~\AA$ \\
EW(Mg~II 2803) & $(1.52 \pm 0.02)$ Å & $< 0.29~\AA$ \\
EW ratio & $1.2 \pm 0.1$ & --- \\
\midrule
$\log (N_\textrm{Mg~II}~/~\textrm{cm}^{-2}) $ & $> 13.8$ & $< 13.1$  \\
log(\NHI~/~cm$^{-2}$)$^\dagger$ & $20.2 \pm 0.1$ & $19.3 \pm 0.10$ \\
$\log (\NHI~/~N_\textrm{Mg~II}) $ & $ < 6.4$ & $ > 6.2$\\
\midrule 
\end{tabular}
\vspace{2mm}
\caption{\footnotesize Equivalent width of the Mg~II transitions and corresponding column densities for different velocity components. The components at $v=-47$ km/s and $v=-164$ km/s are blended and are considered as a single component; the 3-$\sigma$ lower limit on the Mg~II column density is derived from the weaker transition (Mg~II~2803), which is more constraining. For the undetected component at $-640$~km/s, the column density is the average of the values obtained using each of the two transitions. \\
$^\dagger$ from \citet{Srianand2008}.}
\label{tab:mgii}
\end{table}

\begin{table}[tb]
\centering
\renewcommand{\arraystretch}{1.3} 
\setlength{\tabcolsep}{10pt} 
\begin{tabular}{c|c}
\multicolumn{2}{c}{\textbf{\large Mg~I Equivalent Width and Column Density}} \\
\midrule
\midrule
\textbf{Component} & \textbf{-47 km/s} \\
\midrule
EW(Mg~I 2853) & $(0.28 \pm 0.02)$ Å \\
\midrule
log($\NMgI~/~\textrm{cm}^{-2}$) & $12.3 \pm 0.1$\\
log(\NHI~/~cm$^{-2}$)$^\dagger$ & $20.1 \pm 0.1$\\
log(\NHI~/~\NMgI) & $7.8\pm 0.1$ \\
\midrule
\end{tabular}
\vspace{2mm}
\caption{\footnotesize Equivalent width and column density measurements of Mg~I for the component at $v=-47$ km/s. \\
$^\dagger$ from \citet{Srianand2008}.}
\label{tab:mgi}
\end{table}

\begin{table}[tb]
\centering
\renewcommand{\arraystretch}{1.3} 
\setlength{\tabcolsep}{10pt} 
\begin{tabular}{c|c}
\multicolumn{2}{c}{\textbf{\large Fe~II Equivalent Widths and Column Densities}} \\
\midrule
\midrule
\textbf{Component} & \textbf{-47 km/s} \& \textbf{-164 km/s} \\
\midrule
EW(Fe~II 2586) & $(0.51 \pm 0.03)~\AA$ \\
EW(Fe~II 2600) & $(0.99 \pm 0.03)~\AA$ \\
EW ratio & $2.0 \pm 0.1$ \\
\midrule
$\log(\NFeII~/~\textrm{cm}^{-2}) $ & $> 14.0$ \\
log(\NHI~/~cm$^{-2}$)$^\dagger$ & $20.2 \pm 0.1$ \\
$\log (\NHI~/~\NFeII) $ & $ < 6.2$ \\
\midrule 
\end{tabular}
\vspace{2mm}
\caption{\footnotesize Equivalent width of the Fe~II transitions for the blend of the components at $v=-47$ km/s and $v=-164$ km/s. The 3-$\sigma$ lower limit on the Fe~II column density is derived from the weaker transition (Fe~II~$2586$), which is more constraining. \\
$^\dagger$ from \citet{Srianand2008}.}
\label{tab:feii}
\end{table}

\end{appendix}

\end{document}